# Leadership without Leaders?
# Starters and Followers in Online Collective Action

Helen Z. Margetts*, Peter John†, Scott A. Hale*, Stéphane Reissfelder*
*Oxford Internet Institute, University of Oxford
†University College London

Abstract

The Internet has been ascribed a prominent role in collective action, particularly with widespread use of social media. But most mobilisations fail. We investigate the characteristics of those few mobilisations that succeed and hypothesise that the presence of 'starters' with low thresholds for joining will determine whether a mobilisation achieves success, as suggested by threshold models. We use experimental data from public good games to identify personality types associated with willingness to start in collective action. We find a significant association between both extraversion and internal locus of control, and willingness to start, while agreeableness is associated with a tendency to follow. Rounds without at least a minimum level of extraversion among the participants are unlikely to be funded, providing some support for the hypothesis.

## Background: online collective action and leadership

The third millennium has begun with many prominent examples of collective action worldwide, from the uprisings against dictatorships in the so-called 'Arab spring' of 2011, to widespread mobilisation against liberal democratic governments in the aftermath of the financial crash and crisis of 2008. Widespread use of the internet, with its capacity to reduce the costs of participation, dissemination and organisation, has often been cited as a contributing factor. Use of the internet seems to be related to political activism even when controlling for prior social or attitudinal characteristics of Internet users such as age, education and civic duty (Norris, 2009, p. 135) and interest in politics (Borge and Cardenal, 2009). A growing number of scholars argue that the availability of Internet-based platforms challenges long-standing conventional wisdom about the limits and barriers to mass political participation (Lupia and Sin, 2003; Bimber et al., 2006, 2008, 2012; Bennett and Segerberg, 2012, 2013), particularly through the facilitation of huge online gatherings of people who do not know each other, and who have carried out small participatory actions, such as signing an e-petition or raising the profile of a demonstration through endorsement or notification on a social media site. The low cost of these acts would mean that they appear low down on any 'ladder' or scale of political participation (Arnstein, 1969; Almond and Verba, 1961; Parry et al, 1992), but they clearly fall within the category of collective action geared at the securing of public goods. And some large-scale mobilisations of these 'micro-donations' of resources have brought new political issues to public attention and even achieved the ultimate goal of policy change. For example, in the UK a handful of petitions submitted to the No 10 Downing Street e-petitions site obtained over a million signatures,[1] one of which was officially recognised as playing a part in reversing the government's policy plans to introduce road pricing.

Even in this golden age for collective action, however, most attempts to mobilise collective action around public goods fail, while only a few succeed. For every successful demonstration, petition, or sustained collective action for policy change, there are many more campaigns for similar issues that have barely left the ground, never obtaining critical mass or success by any criteria. The collection of transactional data from 8,000 electronic petitions to the UK government (shown in Figure 1 below) shows that 95 per cent fail to attain even the 500 signatures necessary for an official response, a modest criterion for success. Use of the Internet for collective action increases the ratio of unsuccessful to successful collective initiatives. Low start-up costs mean that mobilisations that in offline environments would have failed to get off the ground may achieve some sort of presence, but quickly wither away.

---

[1] The site collected more than 8 million signatures from over 5 million unique email addresses from 2009 to 2011, see http://www.mysociety.org/projects/no10-petitions-website/

So what are the characteristics of the very few mobilisations that succeed? In the offline world, the success of political mobilisation was often ascribed to the existence of effective leaders; people or organisations that overcome the co-ordination problems of an interest group or community by organizing and initiating action (Calvert, 1992). Leaders identify public goods and common goals to pursue; they mobilise collective resources such as time, money and organisation; and they create and reinforce collective trust, group identity and cohesion (Colomer, 2011, p. 55), providing selective incentives for individual acts of participation. However, Internet-based platforms drastically reduce the costs of these activities, particularly co-ordination and dissemination, and minimise the cost of individual participatory acts, diminishing the need for collective trust, group cohesion, or selective incentives. Although there have been some claims that charismatic new 'cyberchiefs' will replace traditional leaders and play an increasingly important role in bringing together 'online tribes' (O'Neil, 2009: 43) with collective goals, the majority of work in this field has focused on the idea that 'formal organisations with structures and incentives are no longer critical' for collective action (Bimber et al., 2012, p. 4-5; Benkler, 2006; Shirky, 2008; Bimber et al., 2005; Bimber et al., 2008, Flanagin et al., 2006). Even those scholars who argue for bringing the relevance of formal organisation back into contemporary conceptualisation of collective action (particularly Bimber et al., 2012), or that online social networks are replacing organisations in analyses of collective action, in a new 'logic of connective action' (Bennett and Segerberg, 2012) see a general shift of agency from leaders and elites to members or individuals; 'the flow of almost all innovation in digital media at present is in the same direction, as power is shifted from centers to ends' (Bimber et al., 2012, p. 20).

In this paper we follow this line of argument in assuming that leaders in the traditional sense are less important online, but focus on one aspect of leadership that we hypothesise will still be important for sustainable collective action: the existence of a number of people who are willing to undertake the (low-cost) action of joining early when there are few signals of the mobilisation's viability or that other people will participate, but who do not necessarily need to be willing to undertake co-ordination or organisation costs, or be well resourced in terms of time or money, or be charismatic. The petition data presented in Figure 1 below provide some evidence for this claim; the number of signatures a petition gets in the first day was among the most important factors in predicting whether the petition will eventually succeed (Hale, Margetts, and Yasseri 2013), suggesting that the existence of a certain number of people willing to start a mobilisation at the point when there has been no media coverage, no time for formal organisation to play much of a role, and no reassuring evidence of other substantive support, are essential to viability. Such individuals may be labelled 'starters', or at the most 'initiators', rather than 'leaders', because they perform none of the other actions associated with leaders of interest groups.

**Figure 1. Petitions submitted to No.10 Downing St. E-petitions Site, Sept. 2009 - May 2010**

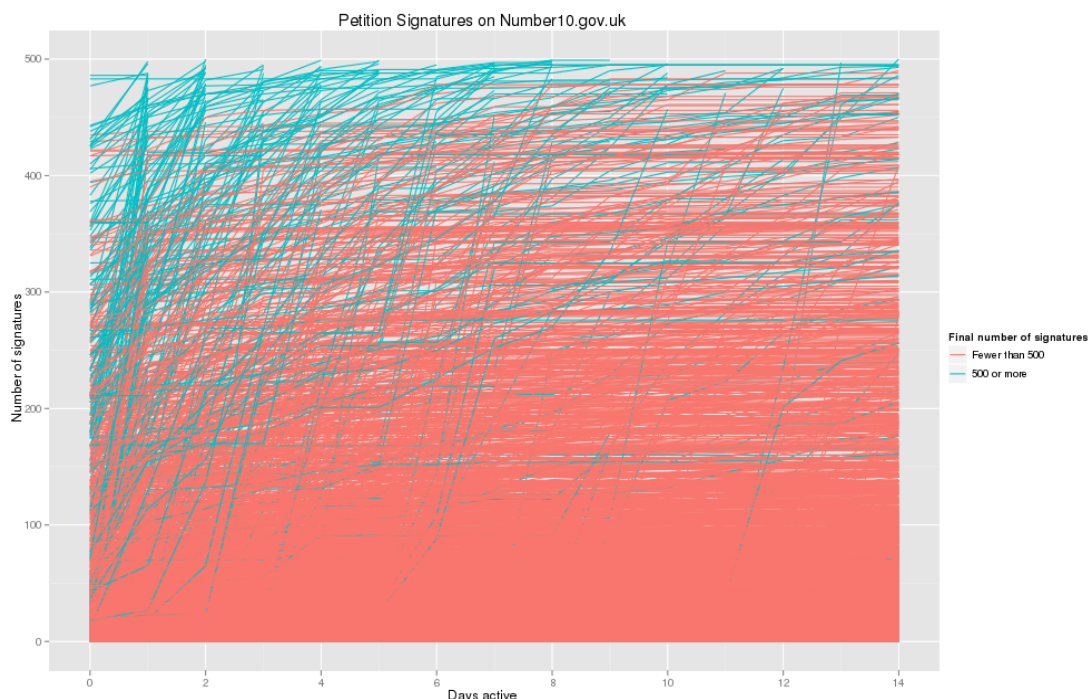

*Source: Data from No.10 Downing Street E-petitions site, Sept 2009 to May 2010 (N= 8,326 petitions)*

Having identified starters as the new leaders, we assume further that they will have distinctive individual-level characteristics which we aim to investigate, just as earlier studies have endeavoured to identify the characteristics of effective leaders in offline settings, for example in terms of charismatic leadership and the leadership styles most likely to engender co-operation (see Judge et al., 2002 for a review). We use experimental data to investigate the characteristics of those people who are likely to join a mobilisation on the critical first day. We use previous work on the characteristics associated with leaders as a starting point for identifying those people most likely to join when there are few other participants, while assuming that not all of them will be relevant to this one activity. We focus on personality as a potential influence on individual behaviour in this regard, following other political scientists who have found personality to be 'comparable to demographics and fundamental political predispositions' (Mondak and Halperin, 2008: 360). The first section below identifies some previous research into individual thresholds for joining mobilisations, while the second section draws together research on leadership that might provide a clue to the personality of those individuals willing to 'start' online mobilisations, allowing us to build three hypotheses. The third section presents the experimental design and results, while the fourth section applies these results to each of the hypotheses developed in the earlier sections and discusses the research and policy implications of the findings.

**Starters, followers and threshold effects**

What characterises starters, the people who are willing to join in the absence of other participants? The clue to identifying differential willingness to join at the start of a mobilisation might come from much earlier work on thresholds in collective behaviour. Schelling (2005) in his 1978 book *Micromotives and Macrobehaviour* developed a model of mobilisation, although not for a political context. He assumed that people have different thresholds for joining a mobilisation—that is, some will join when the number expected to join is low; most will join where the number expected to join is in the middle; and only a few will hold back and join only after the number expected to join is high. He argued that where such thresholds are normally distributed, there will be an S-shaped joining curve for any mobilisation, illustrated in its simplest form in Figure 2 with the number of people that will join plotted against the number that are expected to join. He produces a variety of models from this basic form, showing how sometimes the mobilisation curve will cross the diagonal line, in a tipping point where the number of people expected to participate meets the threshold of most people, and the mobilisation will succeed. In other situations, where the curve does not rise above the line, the mobilisation will fail. Similarly, Granovetter (1978; 1983) identified the concept of threshold as key to the viability of collective behaviour, arguing that the distribution of thresholds was a vital determinant of outcomes.

The threshold model suggests individual behavioural disposition doesn't change before, during and after collective decisions are made; these are contingent dispositions that act on the situation, even if actual behaviour may change. But there is no attempt to identify the origins of these dispositions; neither Schelling nor Granovetter 'consider how individuals happen to have the preferences that they do' (Granovetter, 1978, p. 1421). In contrast, the aim of this research is to investigate empirically the distribution of thresholds, focusing on the characteristics of the group of starters who are willing to join early.

**Figure 2. Schelling's participation curve**

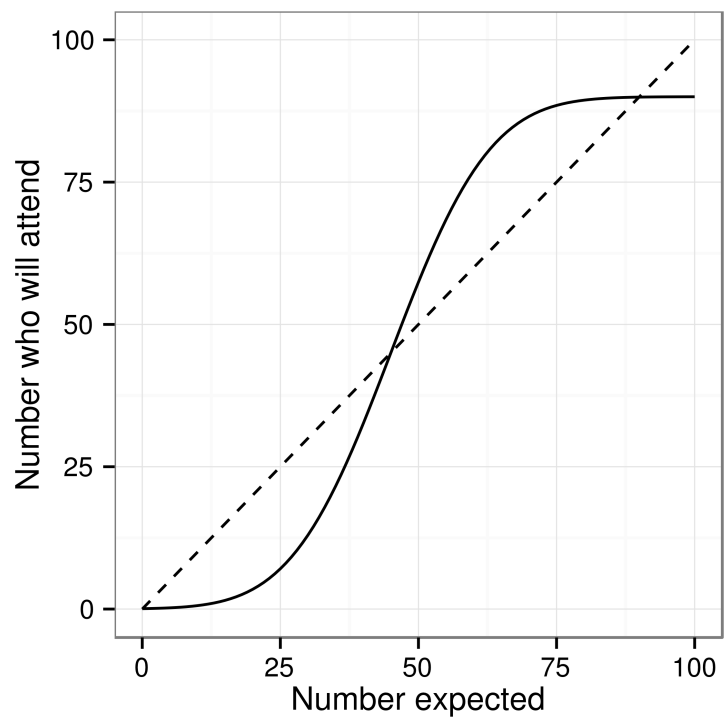

In online mobilisations, Schelling's 'number expected' is replaced with the 'social information' that most electronic interfaces provide, that is, an indication of the number of other participants there have been involved in an e-petition, a mass email campaign, a Facebook group or a collection for charity. Although not all sites provide this information, most do, and social media such as social networking sites or micro-blogging sites have such information built into their design. So although Schelling makes the questionable assumption that the 'Number expected' figure is readily available, it may be that for the first time, his assumptions are easily met. Here, we use our data to test his claims, leading to our first hypothesis:

*$H_1$: When observing propensity to join mobilizations among a pool of potential participants in collective action, we will be able to identify individuals who will consistently join mobilizations early, when there are no signals of viability.*

**Personality and thresholds**

Our aim here is to investigate whether people with particular personalities display a different propensity to join early or late, testing the hypothesis that personality types, as stable attributes of human psychology, are likely to predispose an individual to different timings of collective action. This aim links to recent work in political science on the impact of personality upon other kinds of political behaviour and beliefs (see Mondak and Halperin, 2008; Mondak, 2010; Mondak et al., 2010, 2011; Gerber et al., 2010, 2011; Gallego and Oberski, 2012), following an earlier development in economics and a much longer tradition and central role in psychology. Researchers from different academic disciplines tend to use different categorisations of personality. Here we borrow three typologies: one from economics (social value orientation); one from political science (the 'Big 5' personality traits), and one from psychology (locus of control), allowing us to construct three hypotheses as to what personality types would be most likely to be associated with starting a mobilisation. Previous research on leadership and personality provides no clear picture of how personality traits relate to leadership: 'traditional and contemporary research shows that personality cannot explain leadership' (Andersen, 2006); and 'results of investigations relating personality traits to leadership have been inconsistent and often disappointing' (Judge et al., 2002, p. 765). For this reason, we select three alternative typologies of individual-level difference rather than selecting one typology (for which we could have no unequivocal justification), to maximise our chances of identifying the characteristics of those willing to 'start' online mobilisations.

*Locus of Control:* In the field of organisational and applied psychology, internal locus of control —the extent to which people believe that they have control over their own fate—has been identified as an important personality trait with respect to all kinds of social roles, including leadership. Developed by Rotter (1966) into a measurable scale (later shortened by Carpenter and employed in Carpenter and Seki, 2006), locus of control identifies 'internals', who are those people who believe that they are masters of their fate and perceive a strong link between their actions and consequences, and 'externals', and who do not believe themselves as having direct control of their fate and perceive themselves in a passive role with regard to the external environment. In work settings, internal locus of control has been positively associated with favourable work outcomes and greater job motivation (Ng et al., 2006), high job satisfaction and performance (Judge et al., 2002), and more effective leadership (Spector, 1982). In business studies and management research, people with internal locus of control have been identified as more likely to be leaders and are associated with superior leadership performance (Anderson and Schneier, 1978), although Judge et al. (2002) found locus of control to be a weaker predictor of successful leaders than other personality traits reviewed. In economics, Boone et al. (1999) find that 'internals' in social dilemma situations are more likely than 'externals' to play so as to try to influence the behaviour of other players to achieve their goals, while externals behaved less strategically. With particular relevance for this study, locus of control has been associated with the tendency for people to exert active control over the environment (Ng et al., 2006); propensity to participate in social action (Gore and Rotter, 1963); and political activism (Carlson and Hyde, 1980; Guyton, 1988; Milbraith and Goel, 1977).

$H_{2a}$: *'Starters' have a higher internal locus of control than 'followers'*

*Social value orientation*: In economics, the typology of individual difference most akin to personality that is frequently used is that of social value orientation, a measure of personal values. Social value orientation has been used particularly extensively in laboratory experiments involving public goods and co-operation games (Fischbacher et al., 2001; Keser and van Winden, 2000) and, more recently, field experiments in which subjects are provided with varying levels of information about the participation of others (Frey and Meier, 2004; Shang and Croson, 2009). This work has shown that people 'differ strongly in their contribution preferences' (Fishbacher and Gächter, 2010) and identified categorisations which develop the simple distinction between individualistic and co-operative (see Suleiman and Rapoport, 1992; Weimann, 1994; Fischbacher et al., 2001; Fischbacher and Gächter, 2010) to identify some participants as 'strong free riders', some as 'reciprocators' and some as 'conditional co-operators'. In general, these investigations have focused on what effect people's social value orientations have on their preferences for a leader (de Cremer, 2000), and which social value orientation makes for the best leaders and most effective mobilisations (Gachter et al., 2010), rather than effect of social value orientation on willingness to act as leaders. But a range of research showing that pro-social individuals are more likely to undertake the costs of participation (e.g. Cameron et al, 2006; van Vugt et al, 1995), suggests the following hypothesis with respect to willingness to start such mobilisations:

*$H_{2b}$: 'Starters' will have a more co-operative (pro-social) social value orientation than followers*

*'Big 5' personality traits*: Finally, in the recent revival of interest in personality as a predictor of political behaviour, political scientists have made use of the 'Big 5' personality traits, which have been foundational in psychological studies (Wiggins, 1996; John and Srivastava, 1999: 121) and for which a measurement instrument has been developed for situations where very short measures are needed (Gosling et al., 2003). Psychologists have reached a working consensus that these personality traits can be comprehensively conceptualised and reliably measured (Gerber et al., 2010: 111) and are stable over time (Tickle et al., 2001). When used in political science, they have been shown to be significant predictors of political attitudes and propensity to participate, such as turnout (Gerber et al., 2010; Mondak and Halperin, 2008). The 'Big 5' personality factors comprise openness, conscientiousness, extraversion, agreeableness, and neuroticism. Openness refers to people who are inventive and curious rather than cautious; conscientiousness is about being self-disciplined and efficient rather than easy-going or careless; extraversion is about being outgoing and energetic rather than shy or reserved, and extraverted people have positive emotions and wish to seek stimulation in the company of others; agreeableness describes people who are friendly and compassionate rather than cold or unkind; and emotional stability describes even-temperedness and contrasts with a neurotic temperament and anxiety.

From the findings of this political science research, extraversion emerges as the strongest predictor of participation in collective action, particularly group-oriented tasks and social forms of political activity. Mondak and Halperin (2008) found that extravert personalities were consistently associated with higher propensity to participate in politics, a finding verified by Gerber et al (2011), although only with forms of political activity that involve interacting with others, rather than donation to campaigns, for example. For all the other Big 5 traits, results are mixed. Openness to experience has been positively associated with some participatory acts (Mondak, 2010), and this finding has been replicated with non-U.S. samples (Mondak et al, 2011; Vecchione and Caprara, 2009) but not other US samples (Gerber et al, 2011). Inconsistent effects have also been reported for agreeableness, indicating the importance of context for this variable (Gerber et al, 2011). Mondak and Halperin (2008), Mondak et al (2010) and Gerber et al (2011) found that agreeableness was negatively associated with some types of participation (electoral politics) and positively associated with others (such as attending meetings and signing petitions); Volk et al (2011) found that agreeableness was indicative of greater preference for co-operation; while Vecchione and Caprara (2009) found no results for agreeableness. Conscientiousness has been associated with performance in the workplace (Mondak et al, 2011, 215), but has largely been shown to have no effect on civic engagement; Mondak and Halperin (2008) and Gerber et al (2011) found no consistent pattern for conscientiousness, while Mondak et al (2011) found that the association with participation switched from positive for community engagement to negative for participation in protest activity. Finally, mixed results have been found for the association of emotional stability and political participation. Null results were reported by Parkes and Razavi (2004) and Vecchione and Caprara (2009); Mondak and Halperin (2008) and Gerber et al. (2011) found a positive relationship between emotional stability and some forms of participation, but Mondak et al (2010) found negative associations.

From psychology, research investigating the relationship between leadership and the 'Big 5' personality traits has direct relevance to our interest in people's propensity to 'start' or 'initiate' action, rather than merely to participate. That work looks at two broad categories of leadership: leadership emergence and leadership effectiveness (Lord et al., 1986), the latter being less relevant to the current study. Although much of that work focuses on leaders in the sense of other people's views of the potential leader, rather than the 'starter' criteria that we are interested in here, and some researchers have dismissed trait theory as obsolete in this context (Conger and Kanungo, 1998), two extensive and detailed reviews of leadership research (Lord et al., 1986; Judge et al., 2002) find evidence to suggest that the 'Big 5' typology 'is a fruitful basis for examining the dispositional predictors of leadership' (Judge et al., 2002, p. 773). Of the 'Big 5' characteristics, extraversion in particular has been positively associated to self and peer ratings of leadership (Gough, 1990; Judge et al., 2002). In their meta-analysis of leadership and personality research, Judge et al. (2002, p. 773) found that 'Extraversion emerged as the most consistent correlate of leadership', particularly with respect to leader emergence, although there were some findings for conscientiousness and openness to experience. For these reasons, we test all the 'Big 5' personality traits, while hypothesizing that extraversion will be the most important and the most likely to lead to a positive association.

*$H_{2c}$: 'Starters' will be people scoring highly on the personality trait of 'extraversion', while 'followers' will score lowly on extraversion.*

The testing of these hypotheses (1 and 2a-c) should enable us to identify those personality types associated with low thresholds for joining a mobilisation. Finally, we make a hypothesis with respect to the composition of potential participants. If Schelling is right, then the existence of a sufficient number of starters, people who are willing to contribute early, will be crucial to the survival of a mobilisation. If thresholds are normally distributed in a population, then there will be a small number of people with low thresholds to make the crucial first steps, which then encourage other people with higher thresholds to join. If our previous analysis has identified personality traits that can be associated with threshold, then having at least one or two individuals in a group with those personality traits will be crucial to its success, leading to our third hypothesis:

*$H_3$: There will be a positive association between the success of a mobilisation and the number of individuals with low joining thresholds.*

**Research Design**

To test these hypotheses we use data from a one-shot public goods game laboratory experiment undertaken by the authors (Margetts, et al. 2013). Such an environment offers a viable way to explore the hypotheses detailed above because of the unique capability for theory testing (Morton and Williams 2010, Bardsley et al 2010), in spite of the well-known external validity problems of laboratory experiments when compared with field experiments (Levitt and List 2007a, 2007b). As noted above, the online environment lowers resource barriers to participation (Bimber et al, 2006; 2008; 2012). In such a context, individual level factors such as personality and social value orientation (which are much easier to measure in a laboratory setting where a post-experiment questionnaire may be administered) are likely to come to the fore, in determining why people engage, and contextual factors (which pose particular design and measurement issues in simulated laboratory settings) are less important.[2] So it may be that online participation can be more successfully measured in a laboratory experiment than its offline equivalent. However, it is important to address the external validity challenges of laboratory experiments, doing whatever we can to mitigate them through our research design, as detailed below.

In the experiment, 186 subjects were provided with a number of local 'public goods' scenarios and asked to contribute tokens to the collective good in a series of 28 'rounds', which they played anonymously in small groups (7-10), randomly allocated and changing for each round. All subjects were given 10 tokens at the start of each round and shown a public goods scenario (see appendix for scenarios and instructions to subjects). Each round a provision point was set at 60 per cent of the maximum, total possible contribution of all subjects in the group. If reached, an additional payout would be made. The payoffs for alternative strategies were clearly laid out; subjects would benefit most if they were to free-ride (i.e. contribute no tokens) and the provision point was met; and least if they were to contribute all their tokens and the provision point was not met. Even contributing all of one's tokens but meeting the provision point resulted in a higher payoff than not contributing any tokens but not meeting the provision point. Subjects were told they would be paid for one, randomly-selected round. This design has been shown to impel subjects to treat each round as if it were the only one (Bardsley, 2000). In this paper we report only those rounds where they were provided with social information in real-time about the number of other participations who had contributed tokens; each subject took part in between seven and fourteen of these 'social information' rounds. Each round was planned to last 50 seconds, but if there was activity in the final five second, the time was extended an additional five seconds to avoid deadline effects. We used standard questionnaires (see Appendix A) to collect personality information on the personality characteristics outlined above (locus of control, social value orientation and the 'Big 5' personality traits) after the experiment ended.

We test our hypotheses by examining the relationship between personality and the order in which subjects participated in each round by contributing tokens. We use just those rounds where participants were aware of how many others in their group had contributed and whether they were leading or following at the precise moment where they decided to participate, thereby using the number of other participants as a surrogate for 'number expected' in Schelling's model. We look at the rank ($1^{st}$, $2^{nd}$, $3^{rd}$ and so on) at which individuals contribute, looking for consistent patterns in the rank at which they join mobilisations to identify thresholds and then examining the relationship between personality and threshold. We also examine the relationship between the propensity of a round to get funded, and the distribution of personality characteristics across the people in the group playing that round.

## Results

### Identifying thresholds

First, we tested $H_1$ by looking at the propensity of individual subjects to contribute tokens in the early stages, when there was little or no evidence that others were participating. We examined the minimum, median and mean rank at which subjects participated in a round as their threshold for participation, and the distribution of all three is shown in Figure 3. First, with respect to minimum rank, if an individual subject has a minimum rank of 1, it indicates that they are willing to start a mobilisation with no other participants, whereas if their minimum rank across the 28 rounds is 5, then a higher threshold is suggested. Figure 3a shows the distribution of minimum rank across our subjects, showing that just over half of them were willing to 'start' at some point. This result was quite surprising, suggesting a far from normal distribution of people's willingness to start a mobilisation. To identify threshold, however, we need to identify consistent starters, rather than just people who are willing to start at some point. Therefore, second, we look at median rank, and the results are shown in Figure 3b. Here the figures are more evenly distributed, with a few subjects starting often (and therefore having a median rank of 1 or 2), a few almost always joining only at the end of the mobilisation, and most being somewhere in between. The most common median rank is 3, suggesting people who are most likely to contribute when over 25 per cent (two out of eight, the most common group size) have already done so. Third, Figure 3c shows the results for mean rank, showing again a distribution that appears normal and fails all tests for non-normality.

---

[2] We are grateful to an anonymous referee of an earlier version of this article for this point.

**Figure 3. Distribution of subjects' minimum, median and mean ranks and distribution of the number of times subjects started a round**

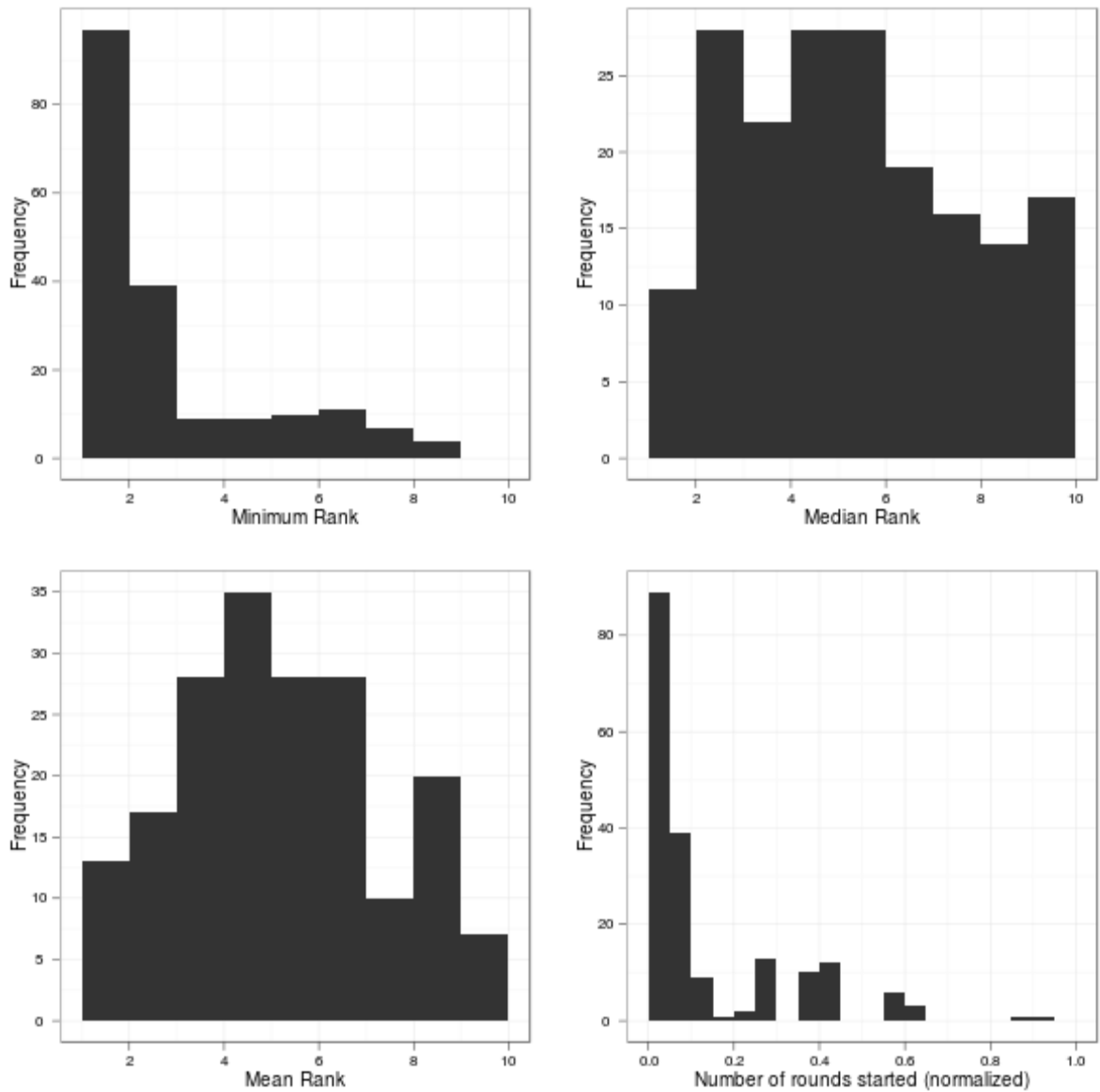

Finally, we construct a variable to count the number of times a subject starts a round (propensity to start) normalised by the number of rounds the subject played. This variable exhibits an extremely skewed distribution (Figure 3d) with just under half of the participants never starting a round, and for whom the 'propensity to start' variable is therefore zero. Of those subjects who do start a round at some point (i.e. who have a value of one or greater and by consequence a minimum rank of one), the distribution is still right-skewed, with most individuals only starting one or two rounds.[3]

[3] Skewness ranges between 1.015 and 1.05 depending on the method used from the three specified in Joanes and Gill (1998).

**Figure 4: Average rank and standard deviation of rank per subject**

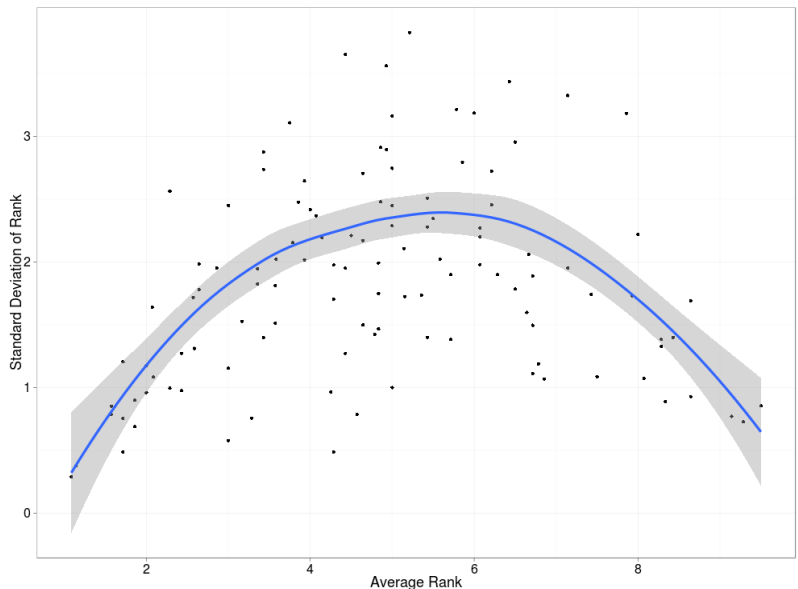

*Note: The best fit line is a locally weighted scatterplot smoothing (LOWESS) line with a 95% confidence interval shaded around the line.*

Each subject played between seven and fourteen rounds in the conditions investigated here. We calculate the standard deviation of each subject's ranks across these rounds as a measure of their consistency. Plotting this standard deviation score against the average rank of each participant (Figure 4) shows that nearly all subjects with a low or high average rank consistently contributed early or late. In contrast, while some subjects with an average rank towards the middle also had a low standard deviation score, other subjects were much less consistent and had higher standard deviation scores. Standard deviations for subjects with a mean rank between 4 and 6 ranged from 0.488 to 3.827 with a mean deviation of 2.276 across subjects. These distributions show that the subjects have heterogeneous propensities to start or join a collective action. It confirms $H_1$ by showing that a few people have consistently low thresholds for joining, and some have consistently high thresholds. There is greater variability in subjects going near the middle (mean standard deviation of 2.276 for subjects with a mean rank between 4 and 6), but early movers and late movers act more consistently.[4]

*Linking thresholds to personality*
The analysis above has identified four possible ways to identify the threshold for individual subjects: Minimum rank; Median rank; Mean rank; and Propensity to Start (a count of number of times contributing first). All four of these measures exhibit variance across subjects, supporting hypothesis $H_{1a}$. Before we could examine the relationship between these threshold measures and personality, we first examined the personality variables in our dataset for collinearity, the results of which are shown in Table 1. These results show that in general the variables are not highly correlated; even the correlations of 0.34 and 0.35 between Rotter score (the variable used to measure internal locus of control) and three of the 'Big 5' variables (Extravert, Open, and Emotionally Stable) are not problematic. Nonetheless, we ran separate regressions for the 'Big 5' and Rotter score when testing our hypotheses and they are shown in separate tables below (see Figures 6a, b).

---

[4] Subjects with a mean rank greater than six have a mean standard deviation of 1.796. Subjects with a mean rank less than or equal to four have a standard deviation of 1.666.

**Table 1. Correlation matrix for personality measures**

| | Extravert | Agreeable | Consci-entious | Emotionally Stable | Open | Rotter Score | Pro-self | Pro-social |
|---|---|---|---|---|---|---|---|---|
| **Extravert** | 1.00 | | | | | | | |
| **Agreeable** | 0.19 | 1.00 | | | | | | |
| **Conscientious** | -0.06 | 0.02 | 1.00 | | | | | |
| **Emotionally Stable** | 0.29 | 0.20 | 0.29 | 1.00 | | | | |
| **Open** | 0.31 | 0.00 | 0.02 | -0.01 | 1.00 | | | |
| **Rotter Score** | 0.34 | 0.05 | 0.20 | 0.34 | 0.35 | 1.00 | | |
| **Pro-self** | -0.18 | 0.10 | 0.06 | 0.01 | 0.04 | -0.08 | 1.00 | |
| **Pro-social** | 0.11 | -0.11 | -0.05 | -0.02 | -0.10 | 0.09 | -0.89 | 1.00 |

We ran tobit regressions to test the association between the threshold measures and the various personality traits discussed above, in order to test hypotheses $H_{2a}$, $H_{2b}$ and $H_{2a}$, and these results are shown in Table 2.[5] We tested for possible associations between demographic variables (income, age and ethnicity) and propensity to start, but none were found to be significant. In all regressions we controlled for the relative importance of the scenario at hand (as asked in the post-experiment questionnaire) for our subjects, as clearly the importance ascribed to an issue will be an important factor in the motivation of an individual to participate.[6]

First, we found locus of control, the personality variable most often associated with leadership in the psychology and management literature to be significantly associated with going early in mobilisations. The locus of control variable was a predictor of median rank, mean rank, minimum rank, and propensity to start (Table 2a) across subjects. So $H_{2a}$ is confirmed, although as in some earlier work (Judge, 2002), locus of control was found to have a weaker effect than some of the other personality traits we investigated.

Second, we found no significant association between threshold and social value orientation, the most consistent predictor of contribution amount and susceptibility to social information in our previous analysis (Margetts, et al. 2013). Neither 'pro-self' nor 'pro-social' individuals were more or less likely to start or follow other subjects. So $H_{2b}$ is rejected and we exclude social value orientation from the rest of the analysis.

Third, we found significant results for two of the 'Big 5' personality traits, identified in recent political science research as important predictors of political behaviour. Across our measures of threshold, we found that 'extraversion' was consistently associated with going early, and 'agreeableness' was consistently associated with going late. Table 2b shows agreeable people as being strongly associated with higher median ranks and higher mean ranks, and lower propensity to start. Extraversion is also an important personality variable. Extravert people have a higher median rank (significant to 0.05), and a higher mean rank (significant to 0.02), and they are significantly more likely to have a high number of starts (significant to 0.001). None of the other 'Big 5' personality traits have a significant effect on these 'threshold' variables. These results confirm Hypothesis 2a with respect to extraversion, with the additional finding for agreeableness, which has a consistently higher coefficient than those for extraversion.

[5] We use double-sided tobit because the dependent variable is censored. Participants cannot give less than zero tokens or more than 10 tokens in any round. Thus the sum of a participant's contributions have a floor and ceiling.
[6] The post-experiment questionnaire included questions on whether the subject agreed with the issue, and how important they considered it to be. Both were significant but as they were correlated, we controlled for importance as it had the most predictive strength.

**Table 2. Tobit regressions showing relationships between threshold measures & personality traits**

**a. Rotter Score**

| | Rank | | | |
|---|---|---|---|---|
| | Average | Median | Minimum | Propensity to start |
| Mean Importance | -3.168* | -3.793* | -5.211* | 0.115 |
| | (1.34) | (1.65) | (2.54) | (0.21) |
| Rotter Score | -2.226* | -3.065** | -5.674** | 0.421** |
| | (0.92) | (1.14) | (1.81) | (0.15) |
| Constant | 8.853*** | 9.673*** | 8.036*** | -0.333 |
| | (1.15) | (1.41) | (2.18) | (0.19) |
| N | 177 | 177 | 177 | 177 |

*p<0.05, ∗∗p<0.01,∗∗∗ p<0.001*

**b. 'Big 5' Personality Traits**

| | Rank | | | |
|---|---|---|---|---|
| | Average | Median | Minimum | Propensity to start |
| Mean Importance | -3.649** | -4.164* | -5.454* | 0.155 |
| | (1.29) | (1.62) | (2.45) | (0.10) |
| Extravert | -1.788** | -2.073** | -3.428** | 0.276** |
| | (0.61) | (0.78) | (1.20) | (0.10) |
| Agreeable | 3.063*** | 2.936** | 5.715*** | -0.606*** |
| | (0.71) | (0.90) | (1.44) | (0.12) |
| Conscientious | -0.154 | -0.421 | 0.665 | -0.033 |
| | (0.79) | (1.00) | (1.60) | (0.12) |
| Emotionally Stable | -0.166 | -0.221 | -0.635 | 0.186 |
| | (0.63) | (0.80) | (1.25) | (0.10) |
| Open | 0.394 | 0.240 | 0.238 | 0.214 |
| | (0.88) | (1.12) | (1.72) | (0.14) |
| Constant | 6.477*** | 7.253*** | 2.224 | -0.086 |
| | (1.27) | (1.61) | (2.43) | (0.20) |
| N | 180 | 180 | 180 | 180 |

*p<0.05, ∗∗p<0.01,∗∗∗ p<0.001*

*Conditions for successful mobilisation*

If successful mobilisations rely on a number of people with low thresholds, and if these people are those with extravert personalities (or high internal locus of control), then it should be that rounds without people with one or other of these personality types will consistently fail, as hypothesised in $H_3$. To test $H_3$ we collapsed data at round level, looking at the maximum, minimum and mean scores for internal locus of control, extraversion and locus of control in each round and the relationship between these variables and whether a round was funded or not; that is, whether the provision point was met.

We ran regressions for all personality variables separately while controlling for the mean importance of all participants on the team to the scenario at hand (as above) and the amount of the first contribution. Basically, the higher the importance ascribed to the issue across the group of participants, the more likely it is to be supported. The amount of the first contribution in a round is also significant, reaffirming results of previous experimental research in economics investigating conditional co-operation in charitable giving (see Shang and Croson, 2009). We control for these two variables in the regressions that follow.

**Table 3: Round level regressions**

|  | Funded | |
| --- | --- | --- |
| First Contribution | 3.306*** | 3.310*** |
|  | (0.73) | (0.77) |
| Mean Importance | 4.051* | 4.450* |
|  | (1.72) | (1.75) |
| Mean Extravert | 5.581* |  |
|  | (2.79) |  |
| Min Extravert |  | 4.691** |
|  |  | (1.47) |
| Constant | -9.856*** | -7.969*** |
|  | (2.22) | (1.53) |
| N | 217 | 217 |

$*p<0.05, **p<0.01, ***p<0.001$

    Among all the personality variables tested, only extraversion was significant in explaining the likelihood of a round being funded. Two measures of the extraversion of the group, the mean and minimum, were significant, as shown in Table 3. That the minimum level of extraversion would be associated with whether a round gets funded is commensurate with our hypothesis. If we picture Schelling's mobilisation curve shown above (Figure 2), it seems that extraversion (or possibly a combination of extraversion and agreeableness) correlates with the 'willingness to join' the movement (the x-axis). Those high in extraversion are so willing to join that they start. Those with more moderate levels of extraversion have a medium threshold to join. Those with low levels of extraversion have a very low willingness to join, and too high a proportion of these individuals hamper the possible success of a movement. A high minimum value for any given round suggests that there are lower numbers of people with higher thresholds, which seems to be associated with a higher likelihood of a round being funded. Mean extraversion was also significant, suggesting again that generally high numbers of extravert people in a group is associated with a higher chance of a round being funded. We did not, however, find similar results for internal locus of control. Similarly, the minimum, maximum and mean levels of agreeableness (and all the other personality variables) were not associated with the likelihood of a round being funded.

**Discussion**
We have identified heterogeneity in our experimental subjects in their willingness to join a collective action at an early stage, when there is little or no evidence that other people will join or that the common goal is likely to be attained. That is, some people were willing to start a mobilisation more than once; some joined mobilisations only during the later stages; and most varied their behaviour across a middle range. We have also identified that for a proportion of subjects in our experiment, the rank (or willingness) at which they join is consistent, and we found most evidence of this consistency at the early and late stages of a mobilisation. That is, we identified some people with consistently low thresholds, and some with consistently high thresholds for joining. These findings confirm our first hypothesis, $H_1$.

We have identified the personality characteristics associated with low or high thresholds, although not for all the personality variables that we investigated in our analysis. Social value orientation, for example, seems to be unrelated to people's propensity to start (rejecting $H_{2b}$), in spite of the strong relationship between this variable and contribution behaviour identified by economists (Gachter et al., 2010). In contrast, internal locus of control is positively associated with a willingness to start, confirming $H_{2a}$ and supporting the findings of research in psychology and management on leadership (Anderson and Schneier, 1978; Spector, 1982; Gore and Rotter, 1963; Carlson and Hyde, 1980; Guyton, 1988; Milbrath and Goel, 1977).

Likewise, we find that extraversion is consistently associated with willingness to start, confirming $H_{2c}$. This confirms evidence from political science that points to a relationship between extraversion and political participation (Mondak and Halperin, 2008; Gerber et al, 2011). We might have expected that in online contexts where participants to not interact socially, such as the one we simulated here, extraversion would be less predictive (as Mondak et al, 2010 and Gerber et al, 2011 found for various 'non-social' types of participation), so the fact that it remained significant throughout our analysis renders this finding more interesting and suggests that this personality trait acts endogenously to predispose individuals to start or initiate action even when there is no indication that others will do so. The positive association we found between the concentration of extraversion in a subject group, and the likelihood of a round getting funded, suggesting that a mobilisation with lower numbers of people with low extraversion (and conversely, higher numbers of people with higher levels of extraversion) has a higher chance of success, reinforces this finding.

We have made the additional finding that agreeableness is positively associated with a tendency to go late, suggesting that agreeable people wait until it is clear that the mobilisation will be funded (that is, that more than half the necessary number have contributed) before joining themselves. This comports with the mixed results of previous research for this personality trait; agreeable people seem drawn to some types of participation and not to other, more conflictual activities; starting an action where there are few other participants could fall into this category, confirming the importance of context and 'personality-dependent characteristics of each participatory act' (Gerber et al, 2011). None of the other 'Big 5' personality variables investigated are significantly associated with threshold, a result that joins the mixed results for these traits in most previous political science research (Mondak, 2010; Mondak and Halperin, 2008; Vecchione and Caprara, 2009; and Gerber et al, 2011).

These findings also support previous research on leadership (Lord et al., 1986; Gough, 1990; Judge et al., 2002) finding that the 'Big 5' typology is a fruitful basis for examining predictors of leadership and that extraversion is the most consistent correlate of leadership (Judge et al., 2002). This previous research focused on leadership in organisations, whereas our findings for extraversion in the different context of collective action are stronger and less equivocal. It may be that the strength of our results derive from extraversion being best suited to the starting element of leadership, and less so to the other activities required for leading in offline contexts, which require other dimensions of personality.

The research design employed here indicates potential for laboratory experiments to investigate the relationship between personality and collective action and joins a growing body of experimental research in political science. As suggested above, online (as opposed to offline) participation may be easier to simulate in a laboratory setting, but the design used here had still to tackle the usual external validity problems of such experiments. The next phase of this research will move towards using an internet-based environment, where online participation may be more accurately simulated, yet the information environment can be more easily manipulated and randomised than in a normal field experiment. However, the challenge of administering a personality questionnaire in a field setting limits the potential for using the internet as a 'field' in this way, suggesting that a micro-labour platform such as Amazon Mechanical Turk could be the most viable solution, with an 'internet-as-laboratory' experimental design (Margetts and Stoker, 2010; Margetts et al, 2011).

**Conclusion**

We started with the puzzle that most online mobilisations do not succeed, while a very few do. We aimed to identify a mechanism that distinguishes the successful ones. Analysis of large-scale transactional data from electronic petitions suggests that the early days of a mobilisation are crucial; success is reliant on a significant number of people being willing to act at this early point, when there is no social information to indicate viability. Those who participate in the early moments set off a chain of reactions—giving an indication of support to others with harder criteria for joining, who in turn, by joining, give a similar indication to those with harder criteria and so on—that results in some mobilisations reaching a critical mass or tipping point. The mechanism that sets off this chain reaction is based on differential thresholds among potential participants; some people are willing to go early, which satisfies the criteria of those whose threshold is a little higher and so on, as in Schelling's model. In online settings, where people can be rapidly notified of new mobilisations through large-scale social networks, and social information is readily available, such a chain of events can happen very quickly; likewise, a movement may fail almost immediately as news of the mobilisation slips down the news feed on social media platforms, and the starting moment is lost. By identifying heterogeneous thresholds, consistent at least at the higher and lower end, we have provided some evidence for this mechanism in action.

In this environment, leadership is the aggregate of many low-cost actions undertaken by those willing to start, rather than the raft of actions and characteristics of the few with which it is normally associated. Of course, the group of starters will usually include at least one leader more in the traditional mould who has taken a higher cost action; for example, the person who sets up a petition and circulates it to close associates in their immediate social networks. But the number of starters needed to get the mobilisation off the ground will be beyond that which could be obtained with strong ties to the initiator alone, but will be attained with weak ties, such as the friend of a friend of a friend on a social networking site, or the retweet of the retweet of a tweet. To obtain the 500 signatures used as a measure of success for the petitions in the analysis in the introduction, the initiator of a petition would have to go beyond their immediate social network; in 2011, the median number of friends on Facebook was 100 (Backstrom, 2011) and the median number of followers on Twitter was 85 (Bakshy et al., 2011). At this point, the raw information of how many signatures have been obtained will be a relatively important piece of social information, rather than the network attachment to the initiator. Many petitions on similar issues fail while one might succeed; it is not the existence of the leader who brings the success, but the existence of a sufficient number of starters with low thresholds, and the readily available and disseminated social information about other participants that draws in the followers with higher thresholds. By providing this social information, Internet-based platforms circumvent the need for other activities traditionally performed by leaders.

If extraversion is a good predictor of low threshold, then the likelihood of a mobilisation obtaining some measure of success will depend in part on the distribution of extraversion in the pool of people who ascribe importance to the issue at hand and are aware of the mobilisation. A possible corollary of this finding is that mobilizations initiated by people with high levels of extraversion are more likely to succeed. Given that we might hypothesise that extravert people tend to select extravert friends (Selfhout et al, 2010), evidence to suggest that extravert people have more friends on social media (Amichai-Haurger and Vinitzky, 2010; Moore and McElroy, 2012) and that an individual is most likely to kick off a mobilisation via their own online social networks, then an extravert's initiative will be immediately disseminated to a pool of people with greater than normal levels of extraversion and, if our findings are correct, low thresholds for joining.

The earlier work by Schelling and Granovetter discussed in the first section assumed that different individuals have different thresholds and based their models on this assumption. They did not investigate why individuals have the thresholds that they do, or what types of people have which types of threshold. In contrast, we have looked at the characteristics of people with low thresholds, and identified a relationship between extraversion (and, to a less clear extent, internal locus of control) and low thresholds. From the evidence reported here, we cannot attempt to make any claims here as to what the distribution of thresholds might be in any given context as our findings are based on experimental data that did not purport to be a representative sample of any population. However, these findings suggest that future research aimed at understanding the distribution of thresholds might include secondary analysis of research into the distribution of extraversion or locus of control in different populations.

These findings also have a wider significance. The argument presented here suggests that contemporary political mobilisations can become viable without leading individuals and organisations to undertake organisation and co-ordination costs, proceeding to critical mass and even achieving the policy or political change at which they are aimed. Simplistically, it could be argued that this is the case with some of the uprisings of the Arab spring, where dictators and their administrative apparatus were toppled without institutions, political parties or embryonic leaders in the wings to undertake the process of rebuilding the state. Single-issue mobilisations that attain critical mass but have no interest group or social enterprise behind them will find it hard to generate sustainable policy change. In this context, understanding the dynamics of these collective actions based on micro-donations of resources becomes increasingly important.

**Acknowledgements**

The experiment reported here forms part of the project, Rediscovering the Civic and Achieving Better Outcomes in Public Policy, funded by the Economic and Social Research Council, Communities and Local Government and the North West Improvement and Efficiency Partnership, grant reference: RES-177-025-0002. The research for this article was also funded in part by the Economic and Social Research Council professorial fellowship held by Helen Margetts entitled 'The Internet, Political Science and Public Policy' [grant number RES-051-27-0331]. We thank our funders.

We also thank those with whom we discussed the ideas and research design of this paper (particularly Josep Colomer), the participants in the seminars, workshops and conference sessions where it was presented in draft form; and the anonymous reviewers of this article.

**Authors**

*Helen Margetts* is Director of the Oxford Internet Institute, where she is Professor of Society and the Internet, and a Fellow of Mansfield College. She is a political scientist specialising in digital government and internet-mediated collective action. She is co-author (with Patrick Dunleavy) of *Digital Era Governance: IT corporations, the state and e-government* (Oxford University Press, 2006).

*Peter John* is Professor of Political Science and Public Policy at University College London. He is known for his work on public policy and agenda-setting. He is interested in how to involve citizens in public policy and management, and in the use of randomized controlled trials, many of which were presented in *Nudge, Nudge, Think, Think: Experimenting with Ways to Change Civic Behaviour* (Bloomsbury 2011).

*Scott A. Hale* is a research assistant and doctoral candidate at the Oxford Internet Institute, University of Oxford. He is interested in developing and applying computer science methods to social science questions in the areas of political science, language, and network analysis.

*Stéphane Reissfelder* is completing a DPhil at Nuffield College, University of Oxford, and was a researcher at the Oxford Internet Institute.